\documentclass[aps,pra,twocolumn,superscriptaddress,amsmath,amssymb,showpacs]{revtex4} 
\usepackage{graphicx}
\usepackage{amsmath}
\usepackage{amsfonts}
\usepackage{amssymb}
\usepackage{hyperref}

\begin{document}

\title{Ground-state phase diagram and critical temperature of two component Bose gases with Rashba spin-orbit coupling}

\author{Zeng-Qiang Yu}

\affiliation{Dipartimento di Fisica, Universit\`{a} di Trento and INO-CNR BEC Center, I-38123 Povo, Italy}

\begin{abstract}

Ground-state phase diagram of two-component Bose gases with Rashba spin-orbit coupling is determined via a variational approach. A phase in which the fully polarized condensate occupies zero momentum is identified.  This zero-momentum phase competes with the spin density wave phase when interspecies interaction is stronger than intraspecies interaction, and the former one is always the ground state for weak spin-orbit coupling. When the energies of these two phases are close, there is a phase separation between them. At finite temperature, such a zero-momentum condensation can be induced by a ferromagnetic phase transition in normal state. The spontaneous spin polarization removes the degeneracy of quasiparticles' energy minima, and consequently the modified density of state accommodates a Bose condensation to appear below a critical temperature.

\end{abstract}

\maketitle

The experimental progress of spin-orbit (SO)-coupled quantum gases has generated great interest recently \cite{NIST,ZhangJing,MIT,ChenShuai}. Unlike the SO coupling extensively studied in solid matters, the synthetic SO coupling in dilute cold atomic gases can be realized in feasible configurations with a tunable coupling strength \cite{Review1,Review3}.  Moreover, it provides a unique platform to study the rich physics of SO effects in a bosonic system \cite{Galitski,Zhai,Ho,half-vortex,LiYun,Ozawa}, which have not been explored before. One particular interesting configuration is Rashba SO coupling \cite{Ruseckas,Dalibard-Spielman}. In free space, since the minima of the single-particle spectrum have infinite degeneracy, the ground state of a Rashba SO-coupled Bose gas is in fact determined by the details of interactions.
Two different Bose-Einstein condensation (BEC) phases, i.\,e., the plane-wave (PW) phase and the spin density wave (SDW) phase (also referred to as the stripe phase), have been identified in a previous study \cite{Zhai}. In a harmonic trap, exotic half-vortex states appear for suitable interaction parameters \cite{half-vortex}. At finite temperature, although a condensate can not survive in a noninteracting system because of the constant density of state (DoS) in the low-energy limit, the BEC state is indeed energetically favored in the presence of repulsive interactions \cite{Ozawa}. In this sense, interactions do help the emergence of condensation.

Previously, many theoretical works focus on the unique degeneracy of the single-particle spectrum. However, in a many-body system, the condensate wavefunction is {\em not necessarily} constructed by the single-particle states with minimal energy. In this work, we present a different ground-state phase diagram as summarized in Fig.~1. In addition to the PW phase and the SDW phase, a fully polarized phase in which the condensate occupies zero momentum (ZM) is identified, which is in strong contrast to single-particle physics. At finite temperature, the emergence of ZM condensation can be induced by a ferromagnetic phase transition in the normal state. The spontaneous spin polarization generates an effective Zeeman field, which significantly modifies the spectrum and the DoS of quasiparticles. Consequently, a finite BEC temperature is retained.

\begin{figure}
\includegraphics[width=7.4cm]{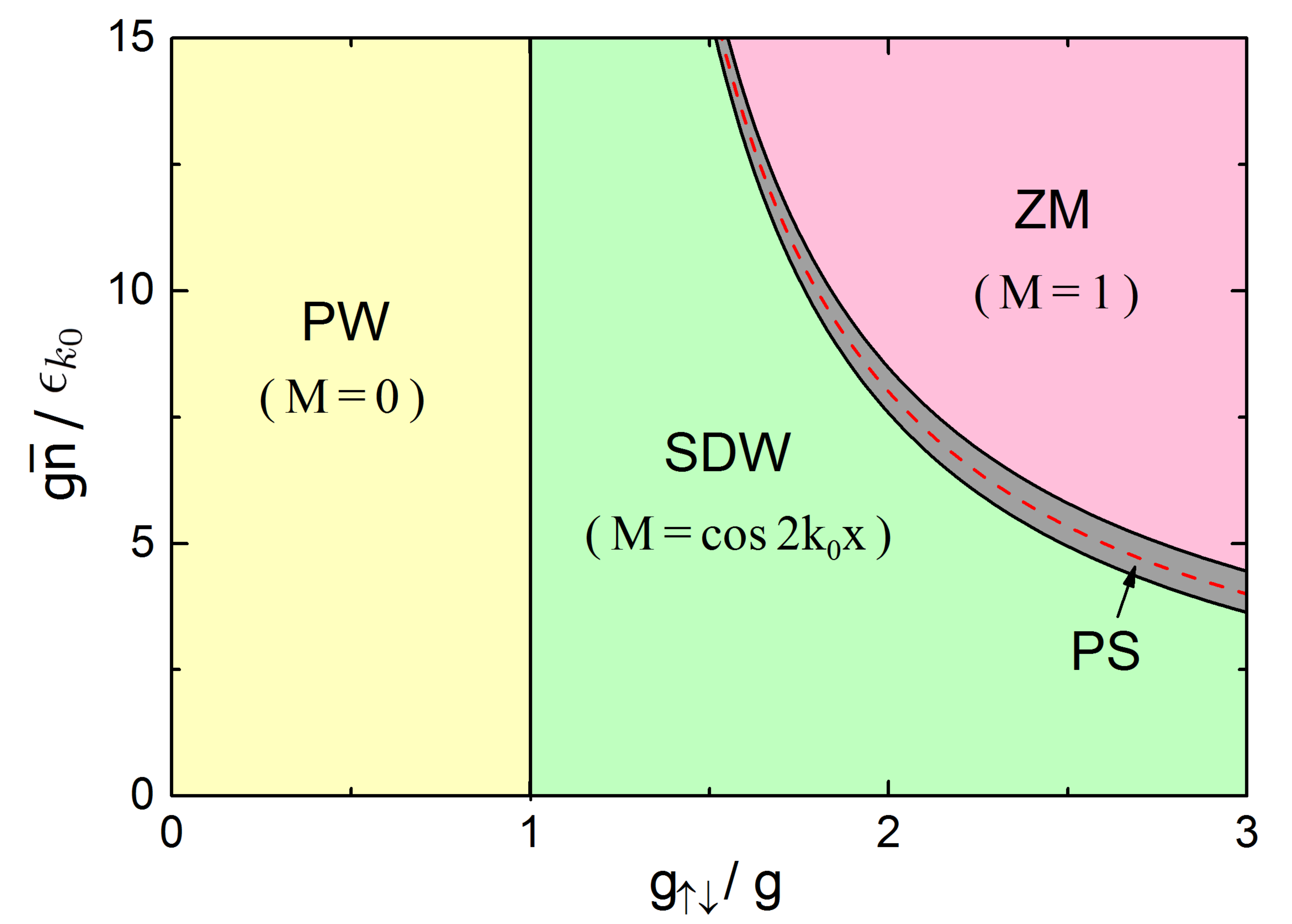}
\caption{(Color online) Ground-state phase diagram of Rashba SO-coupled Bose gases. The three BEC phases, i.\,e., PW phase, SDW phase, and ZM phase, have distinct magnetic properties as indicated by the spin polarization $M$. There is a phase separation (PS) in the shadow region. }\label{fig-1}
\end{figure}

We start from a single-particle Hamiltonian with Rashba SO coupling $(\hbar=1)$,
\begin{equation}
  H_{\rm SO}= \frac{{\bf p}^2}{2m} + \frac{k_0}{m} (p_x\sigma_x+p_y\sigma_y),
\end{equation}
where  $k_0$ is SO coupling strength, and $\sigma_{x,y}$ are Pauli matrices. The eigenstates of this Hamiltonian construct two helicity branches with dispersion $\varepsilon_{\bf p,\pm}=\epsilon_{p} \pm k_0p_\perp/m$. Here, $\epsilon_{p}=p^2/(2m)$, and quantum number helicity $\alpha=\pm$ denotes spin parallel or antiparallel to the in-plane momentum ${\bf p}_\perp$. The minima of the energy spectrum are located at the ring in momentum space with $p_\perp=k_0$. For a noninteracting Bose gas, the ground state has infinite degeneracy, and the condensation wavefunction could be an arbitrary superposition of the single-particle states with the momenta on the degenerate ring.

For a dilute system, interactions between atoms can be described by contact potentials,
\begin{equation}
  \mathcal{\hat H}_{\rm int} = \frac{1}{2}\sum_{\sigma=\uparrow,\downarrow}\int d{\bf r}\, g_{\sigma\sigma'}\hat\psi_\sigma^\dag\hat\psi_{\sigma'}^\dag \hat\psi_{\sigma'}\hat\psi_\sigma
\end{equation}
with $\hat\psi_\sigma$ being the field operator and $g_{\uparrow\downarrow}=g_{\downarrow\uparrow}$. For simplicity, we consider $g_{\uparrow\uparrow}=g_{\downarrow\downarrow}=g$ in this work \cite{note5} and assume all the interactions are repulsive. At the mean-field level, the interaction energy for a BEC state is given by
\begin{equation} \label{E_int}
  \mathcal{E}_{\rm int}  = \frac{1}{4}\int d{\bf r}\,\left[(g+g_{\uparrow\downarrow})n^2({\bf r}) - \delta g\,\bar n^2 M^2({\bf r})\right],
\end{equation}
where $\delta g\equiv g_{\uparrow\downarrow}-g$, $n=n_\uparrow+n_\downarrow$, and $M=(n_\uparrow-n_\downarrow)/\bar n$ are local density and local spin polarization respectively, and $\bar n=N/V$ is average density.

Even without a serious calculation, a qualitative predication can be made from Eq.~(\ref{E_int}). Considering the $\delta g>0$ case, interactions prefer either a fully polarized state or a SDW state to avoid the interspecies repulsion, and the former one has a lower interaction energy. On the other hand, kinetic energy is controlled by the parameter $k_0$. Therefore, when SO coupling is weak enough, kinetic energy can be safely ignored, and the ground state is a fully polarized phase. This argument actually explains an important feature of the phase diagram in Fig.~1.

T ground-state phase diagram can be determined quantitatively via a variational approach. For a state with uniform density and polarization, the condensate wavefunction is generally written as
\begin{equation} \label{PW_wavefun}
   \varphi = \sqrt{\bar n} \begin{pmatrix} \cos\gamma  \\ -\sin\gamma\, e^{i\phi_{{\bf k}}} \end{pmatrix} e^{i{\bf k}\cdot {\bf r}},
\end{equation}
where $\bf k$ is the condensation momentum, $\phi_{\bf k}=\arg(k_x+ik_y)$ for $k\neq 0$ and an arbitrary number for $k=0$, and the parameter $\gamma\in [0, \pi/2]$ determines the spin polarization. The variation energy is readily obtained as
$  \mathcal{E}=N \big(\epsilon_k +\frac{1}{2}g\bar n - \frac{k_0}{ m}k_\perp \sin2\gamma  + \frac{1}{ 4}\delta g\,\bar n\sin^22\gamma \big). $
By minimizing this energy with respect to parameter $\gamma$ and  momentum $\bf k$, we find that when $\delta g\,\bar n<4\epsilon_{k_0}$, an  unpolarized condensate with $(k_\perp,k_z)=(k_0,0)$ is preferred, and the wavefunction in Eq.~(\ref{PW_wavefun}) corresponds to the single-particle energy minima. This state is called the PW phase, whose energy is given by,
\begin{equation}
  \mathcal{E}_{\rm PW} = N\left[ \frac{1}{4}(g_{\uparrow\downarrow}+g)\bar n-\epsilon_{k_0}  \right]. \label{PW_energy}
\end{equation}
When $\delta g\,\bar n>4\epsilon_{k_0}$, the fully polarized condensate which occupies zero momentum is energetically favored, i.\,e., ${\bf k}=0$, $M=\cos2\gamma=\pm 1$. The energy of this ZM phase is given by
\begin{equation}
  \mathcal{E}_{\rm ZM}=\frac{1}{2}Ng\bar n. \label{ZM_energy}
\end{equation}

We note that condensation at ZM is a pure interaction consequence, which is in strong contrast to the single-particle picture. In an equivalent interpretation, the spontaneous spin polarization generates an internal effective Zeeman field $h_{\rm eff}= \delta g\,\bar n M/2$, which modifies the energy minimum. As discussed later, a similar effect also appears in the normal state above BEC temperature.

For a BEC state that breaks translational symmetry, condensate wavefunction can be written as $({\bf k}\neq 0)$
\begin{equation*}
   \varphi =  \sqrt{\bar n} \bigg[ c \begin{pmatrix} \cos\mspace{-2mu}\gamma  \\ -\!\sin\!\gamma\mspace{2mu}  e^{i\phi_{{\bf k}}} \end{pmatrix} e^{i{\bf k}\cdot {\bf r}}  + c'\mspace{-1mu} \begin{pmatrix} \cos\mspace{-2mu}\gamma'  \\ -\!\sin\!\gamma'  e^{i\phi_{{-\mspace{-2mu}\bf k}}} \end{pmatrix} e^{-i{\bf k}\cdot {\bf r}} \bigg],
\end{equation*}
with $|c|^2+|c'|^2=1$ and $\gamma,\gamma'\in [0,\pi/2]$. The ansatz of the above variational wavefunction is a minimal model to accommodate with spin and/or density spacial modulation.
At the mean-field level, only kinetic energy depends on momentum $\bf k$, and a straightforward variation yields $k_z=0$ and
$  k_\perp= k_0 \big(|c|^2\sin2\gamma + |c'|^2\sin2\gamma'\big)$.
Thus, total energy $\mathcal{E}$ can be expressed as a function of $|c|^2, |c'|^2, \gamma$, and $\gamma'$. Although an analytic solution to this variational problem is difficult, numerical energy minimization is straightforward by exhausting all configurations in parameter space. We find that when a non uniform solution has a lower energy than the PW phase and the ZM phase, condensate wavefunction always takes a SDW form:
\begin{equation}
  \varphi = \frac{\sqrt{\bar n}}{2} \bigg[ e^{i\eta} \begin{pmatrix} 1  \\ - e^{i\phi_{{\bf k}}} \end{pmatrix} e^{i {\bf k\cdot r}} + \begin{pmatrix} 1  \\ - e^{i\phi_{{-\bf k}}} \end{pmatrix} e^{-i{\bf k\cdot r}} \bigg], \label{SDW_wf}
\end{equation}
where $(k_\perp,k_z)=(k_0,0)$, and the relative phase $\eta$ determines the node position of the spin density modulation $M({\bf r})=\cos(2{\bf k}\cdot {\bf r}+ \eta)$. The SDW condensation described by Eq.~(\ref{SDW_wf}) was first discussed in Ref.~\cite{Zhai}.

The energy of the SDW phase is given by
\begin{equation}
  \mathcal{E}_{\rm SDW}=N\left[\frac{3g+g_{\uparrow\downarrow}}{8}\bar n-\epsilon_{k_0}\right]. \label{SDW_energy}
\end{equation}
By comparing this energy with Eqs.~(\ref{PW_energy}) and (\ref{ZM_energy}), we obtain the ground-state phase diagram as shown in Fig.~1. When $\delta g<0$, the PW phase is energetically favored; when $\delta g>0$, the ground state is either in SDW phase or ZM phase. The competition between these two phases is controlled by SO coupling strength $k_0$, and a first-order transition takes place at $\delta g\,\bar n=8\epsilon_{k_0}$ (indicated by the dashed-line in Fig.~1). When SO coupling is weak enough, the ZM phase is always the ground state (for $\delta g>0$) \cite{note0}.

Up to now, we assume the ground state is homogeneous. Since the condition for the transition between the SDW phase and ZM phase is density dependent, a phase separation (PS) may appear. The boundaries of the PS region are determined by the balance conditions for chemical potential $\mu=\partial \mathcal{E}/\partial N$ and pressure $P=(\mu N-\mathcal{E})/V$ of these two phases,
\begin{equation}
  \mu_{\rm SDW}(n_1)= \mu_{\rm ZM}(n_2), \quad \;
  P_{\rm SDW}(n_1) = P_{\rm ZM}(n_2).
\end{equation}
From the equation of state in Eqs.~(\ref{ZM_energy}) and (\ref{SDW_energy}), we find
\begin{eqnarray}
  n_1 &=& \frac{4\epsilon_{k_0}}{3g+g_{\uparrow\downarrow}-2\sqrt{g(3g+g_{\uparrow\downarrow})}}, \\
  n_2 &=& \frac{2\epsilon_{k_0}}{\sqrt{g(3g+g_{\uparrow\downarrow})}-2g}.
\end{eqnarray}
For the density $\bar n\in (n_1,n_2)$, a mixture of the ZM phase and the SDW phase has a lower energy than in the homogeneous case, and thus a phase separation takes place.
The PS region is shown as the shadowed area in Fig.~1. When $g\bar n\gg \epsilon_{k_0}$ , the PS region becomes very narrow, and the boundaries approach the first-order transition line $\delta g\,\bar n=8\epsilon_{k_0}$.

Beyond the mean-field level, one can use Bogoliubov theory to take account of fluctuations. Previously, elementary excitations of  PW phase and SDW phase have been studied \cite{Ozawa,Sarma,HanCuiLiao}. Here, we focus on the ZM phase.

\begin{figure}
\includegraphics[width=6.6cm]{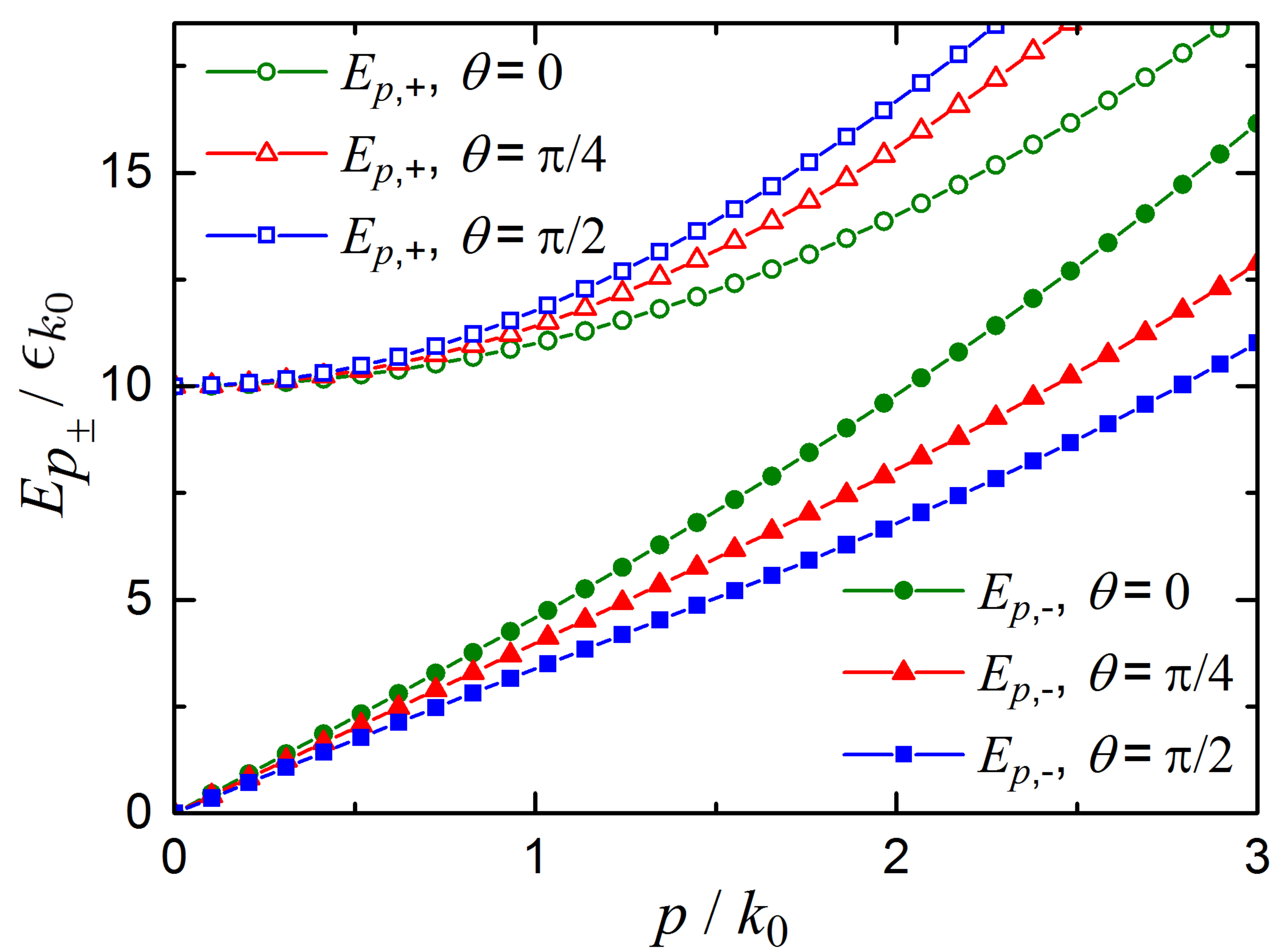}
\caption{(Color online) Elementary excitations of the ZM phase for $g_{\uparrow\downarrow}\bar n=2g\bar n=20\epsilon_{k_0}$. The spectrum has a rotation symmetry in the $p_x$-$p_y$ plane and a reflection symmetry $E_p(\theta)=E_p(\pi-\theta)$.}\label{fig-Ep}
\end{figure}

Assuming a pure spin-up condensate, the Bogoliubov Hamiltonian for the ZM phase is given by
\begin{equation}
  \mathcal{\hat K}_{\rm ZM}=\sum_{p_x>0}\left[\hat\Psi_{\bf p}^\dag \mathbb{K} \hat\Psi_{\bf p} -\big(2\epsilon_p+g_{\uparrow\downarrow}\bar n\big) \right], \label{Bogoliubov}
\end{equation}
with $\hat\Psi_{\bf p}^\dag=(\hat\psi_{\bf p,\uparrow}^\dag,\hat\psi_{\bf p,\downarrow}^\dag,\hat\psi_{-\bf p,\uparrow},\hat\psi_{-\bf p,\downarrow})$ and
\begin{equation}
  \mathbb{K} = \begin{pmatrix} \epsilon_p+g \bar n & \frac{k_0}{ m}p_\perp e^{-i\phi_{\bf p}} & g\bar n & 0 \\
  \frac{k_0}{m}p_\perp e^{i\phi_{\bf p}} & \epsilon_p+\delta g\, \bar n & 0 & 0 \\
  g\bar n & 0 &  \epsilon_p+g\bar n & \frac{k_0}{ m}p_\perp e^{-i\phi_{-\bf p}} \\
  0 & 0 & \frac{k_0}{ m}p_\perp e^{i\phi_{-\bf p}} & \epsilon_p+ \delta g\,\bar n
  \end{pmatrix}. \nonumber
\end{equation}
By diagonalizing Eq.~(\ref{Bogoliubov}), two branches of elementary excitations $E_{\bf p,\pm}$ are obtained. Some examples of $E_{\bf p,\pm}$ along different directions are shown in Fig.~\ref{fig-Ep}. Within the ZM region of the phase diagram, we find the spectra are always positive. This result disagrees with those of previous works \cite{Zhai,Wu_Biao}, which claim the zero-momentum condensate always suffers a dynamical instability \cite{note4}.

The higher excitation branch $E_{\bf p,+}$ possesses an energy gap $\delta g\, \bar n$, which is just the energy cost to flip a spin to overcome the effective Zeeman field. The lower branch $E_{\bf p,-}$ is gapless with a phonon dispersion in the long wave length limit, $\lim_{p\rightarrow 0}E_{\bf p,-}=c p$. The anisotropic sound velocity $c$ depends on direction angle $\theta=\arccos(p_z/p)$,
\begin{equation}
  c(\theta)=\sqrt{\frac{g\bar n}{ m}\left[1-\frac{4\epsilon_{k_0}\sin^2\theta}{(g_{\uparrow\downarrow}-g)\bar n}\right]}, \label{sound}
\end{equation}
and it reaches a minimum value when the phonon mode propagates in the $p_x$-$p_y$ plane.
In the $k_0\rightarrow 0$  limit, Eq.~(\ref{sound}) recovers the usual sound velocity of a single-component BEC.

Now, we consider the normal state without a Bose condensation.

In a noninteracting system, the equation for total density at  temperature $T$ is given by $(k_{\rm B}=1)$
\begin{equation}
  \bar n=\frac{1}{V}\sum_{{\bf p},\alpha=\pm}f(\varepsilon_{\bf p,\alpha})=\int d\varepsilon\, D(\varepsilon)f(\varepsilon), \label{density_eqn}
\end{equation}
where $f(\varepsilon)=\big[e^{(\varepsilon-\mu)/T}-1\big]^{-1}$ is Bose distribution function, and $D(\varepsilon)=\frac{1}{V}\sum_{{\bf p},\alpha}\delta(\varepsilon-\varepsilon_{\bf p,\alpha})$ is the DoS of single-particle. Due to the degeneracy of energy minima, $D(\varepsilon)$ is a constant in the low-energy limit \cite{Vijay}, and such a DoS prevents a condensate surviving at finite temperature.

In the presence of interactions, Eq.~(\ref{density_eqn}) is still valid at the mean-field level with $\varepsilon_{\bf p,\pm}$ and $D(\varepsilon)$ replaced by the dispersion and the DoS of quasiparticles respectively.
If Hartree-Fock energy only makes a constant shift in the quasiparticle spectrum, the DoS remains the same as in the noninteracting case, and the normal state is not kinetically forbidden at any finite temperature \cite{Ozawa}.

The situation is different once a spontaneous magnetization appears. The mean-field Hamiltonian for a normal state with a possible spin polarization $M$ is given by
\begin{align*}
  \mathcal{\hat H}_{\rm MF} &=\, \sum_{\bf p}\bigg[\sum_{\sigma=\uparrow\downarrow} (\epsilon_p+2gn_\sigma+g_{\uparrow\downarrow}n_{\bar\sigma})\hat\psi_{\bf p,\sigma}^\dag \hat\psi_{\bf p,\sigma}  \nonumber \\
  & \qquad  + \Big(\frac{k_0}{m}p_\perp e^{i\phi_{\bf p}} \hat\psi_{\bf p,\downarrow}^\dag \hat\psi_{\bf p,\uparrow} + h.c.\Big) \bigg]-\mathcal{E}_{\rm int}^0, \nonumber
\end{align*}
with $\mathcal{E}_{\rm int}^0=V(gn_\uparrow^2 + gn_\downarrow^2 + g_{\uparrow\downarrow}n_\uparrow n_\downarrow )$ and $\bar\sigma$ the opposite spin to $\sigma$. The quasiparticle energy spectrum $\varepsilon_{\bf p,\pm}=\epsilon_p+(2g+g')\bar n \pm \sqrt{(g'\bar n M)^2+(p_\perp k_0/m)^2}$ has a different structure from the noninteracting case for $M\neq 0$, where the interaction parameter $g'\equiv g_{\uparrow\downarrow}/2-g$ is defined.

Spin polarization $M$ should minimize the free energy
\begin{equation}
  \mathcal{F}_{\rm N}=\mu N +T\sum_{\bf p,\alpha=\pm}\ln \big(1-e^{(\mu-\varepsilon_{\bf p,\alpha})/T}\big)-\mathcal{E}_{\rm int}^0,
\end{equation}
and the stationary point condition $\partial \mathcal{F}_{\rm N}/\partial M=0$ can be explicitly written as
\begin{equation}
   M=\frac{1}{V}\sum_{\bf p}\frac{ g' M\big[ f(\varepsilon_{\bf p,-})-f(\varepsilon_{\bf p,+})\big] }{\sqrt{(g'\bar nM)^2+(p_\perp k_0/m)^2} }, \label{M_eqn}
\end{equation}
which is equivalent to the self-consistency equation $M=\frac{1}{N}\sum_{\bf p}\big(\langle \hat\psi_{\bf p,\uparrow}^\dag \hat\psi_{\bf p,\uparrow}\rangle - \langle \hat\psi_{\bf p,\downarrow}^\dag \hat\psi_{\bf p,\downarrow}\rangle\big)$. For a given temperature, $M$ and $\mu$ can be solved from Eqs.~(\ref{density_eqn}) and (\ref{M_eqn}).

\begin{figure}
\includegraphics[width=8.6cm]{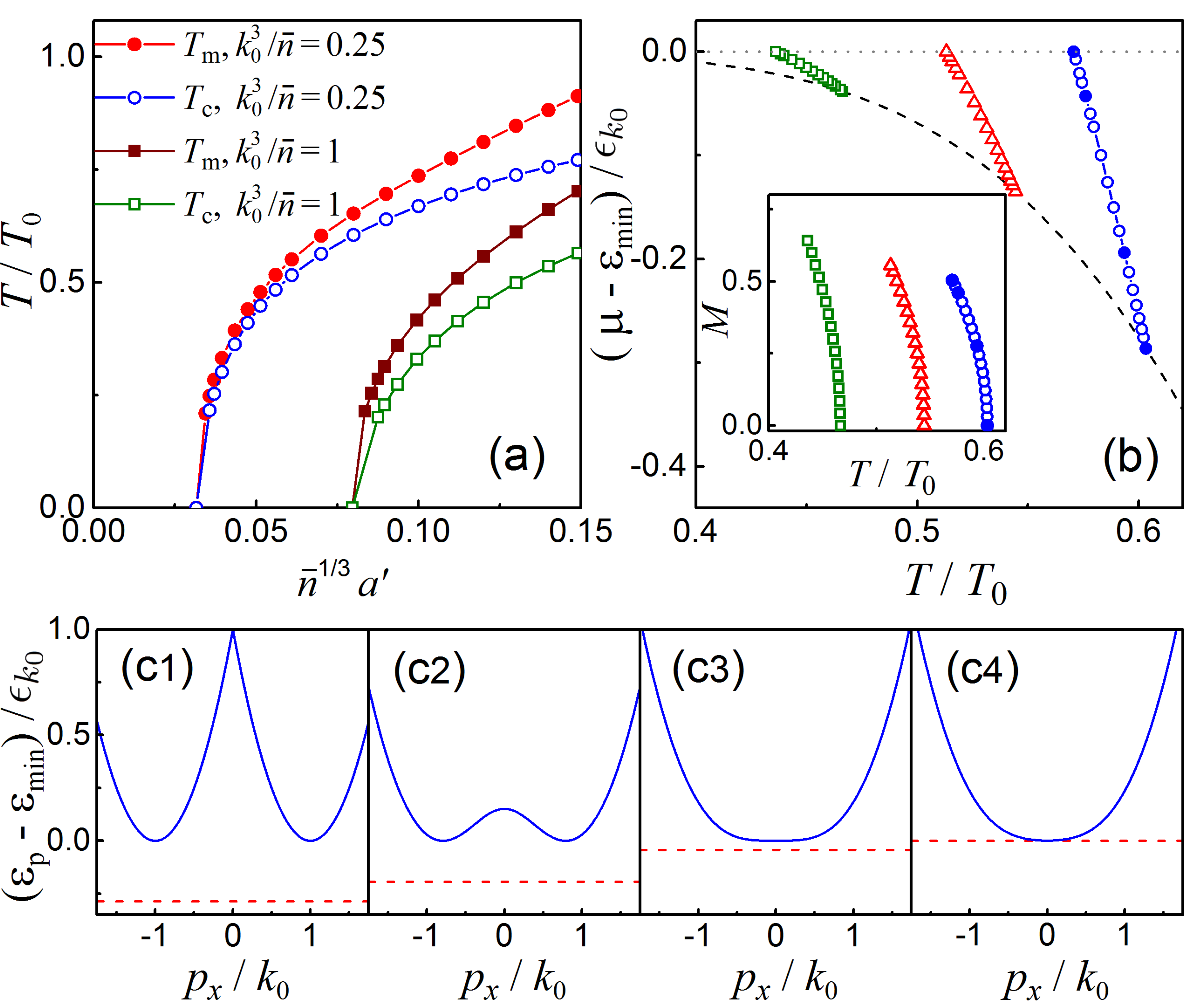}
\caption{(Color online) (a) Ferromagnetic transition temperature $T_{\rm m}$ and condensation temperature $T_{\rm c}$ as functions of  $\bar n^{1/3}a'$. (b) Temperature dependence of chemical potential $\mu$ for $\bar n^{1/3}a'=$ 0.05 ($\square$), 0.06($\vartriangle$), and 0.07($\circ$) with fixed $k_0^3=0.25\bar n$, and the dashed line is for the  noninteracting case. The inset is the corresponding spin polarization $M$ as a function of $T$. (c1)-(c4) Energy spectrum of quasiparticles $\varepsilon_{\bf p,-}$ for the data labeled by $\bullet$ in panel (b), and the dashed lines indicate the value of $\mu-\varepsilon_{\rm min}$ for each case. Here, $T_0\equiv 2\pi\big[\bar n/\zeta(\frac{3}{2})\big]^{2/3}/m$.}\label{fig-Tc}
\end{figure}

For $g'\bar n<2\epsilon_{k_0}$, Eq.~(\ref{M_eqn}) only has a trivial solution $M=0$ at any temperature, which means the normal state is always unpolarized. For $g'\bar n>2\epsilon_{k_0}$, a finite spin magnetization is energetically favored below a ferromagnetic transition temperature $T_{\rm m}$, which is determined by
\begin{equation}
  \frac{1}{g'} = \frac{1}{ V}\sum_{\bf p}\frac{ m}{ p_\perp k_0}\big[ f(\varepsilon_{\bf p,-})-f(\varepsilon_{\bf p,+})\big],
\end{equation}
with $M=0$. In Fig.~\ref{fig-Tc} (a), $T_{\rm m}$ is plotted as a function of $\bar n^{1/3}a'$ for different $k_0^3/\bar n$, with  $a'\equiv g'm/(4\pi)$. One can see that $T_{\rm m}$ increases rapidly when $g'\bar n$ beyond the threshold $2\epsilon_{k_0}$.  For small $k_0^3/\bar n$, the phase transition takes place in the weak interacting regime $(\bar n^{1/3}a'\ll 1)$, where the validity of mean-field treatment can be justified. We note that similar ferromagnetic phase transitions also happen in the normal state of usual spin-1 bosons \cite{Gu} and  spin-$\frac{1}{2}$ bosons \cite{Ozawa_FM} without SO coupling.

Below $T_{\rm m}$,  $M$ increases very quickly as shown in the inset of Fig.~\ref{fig-Tc} (b). The spin polarization generates an effective internal Zeeman field $h_{\rm eff}=g'\bar nM$, which modifies both the spectrum and the DoS of quasiparticles. For $M<2\epsilon_{k_0}/(g'\bar n)$, quasiparticle energy minima still construct a degenerate momentum ring (with a reduced radius), and DoS remains a constant in low energy limit. However, when $M>2\epsilon_{k_0}/(g'\bar n)$, i.\,e., $h_{\rm eff}>2\epsilon_{k_0}$, zero momentum becomes the solo minimum of the quasi-particle spectrum, and the DoS vanishes in low energy limit \cite{note1}. Such a DoS accommodates a condensate to appear, when chemical potential reaches the bottom of quasi-particle spectrum $\varepsilon_{\rm min}$ [see Figs.~\ref{fig-Tc}(c1)-\ref{fig-Tc}(c4)]. Below the condensation temperature $T_{\rm c}$, the normal state does not exist. As shown in Fig.~\ref{fig-Tc}(a), an experiment-achievable $T_{\rm c}$ can be realized  for a reasonable interaction parameter $\bar n^{1/3}a'$.

In contrast to the noninteracting case, the finite BEC temperature obtained here is due to the interaction effect.
We emphasize that the emergence of ZM condensation is induced by a ferromagnetic transition in the normal state, and the critical temperature $T_{\rm c}$ is naturally derived according to Einstein's original idea, which is quite different from the energetic scenario proposed in Ref \cite{Ozawa}. On the other hand, for $g'\bar n<2\epsilon_{k_0}$, the condensation transition must be determined energetically (at the  mean-field level), because normal state the is not kinetically forbidden at any finite temperature. In the $T\mspace{-3mu}\rightarrow\mspace{-3mu} 0$ limit, we find the free energy of the unpolarized normal state for $g'\bar n<2\epsilon_{k_0}$is given by \cite{note2},
$\mathcal{F}_{\rm N}(T\rightarrow 0) = N\big[(2g+g_{\uparrow\downarrow})\bar n/4 - \epsilon_{k_0}\big], $
which is indeed higher than the energy of the BEC state given by Eqs.~(\ref{PW_energy}), (\ref{ZM_energy}), and (\ref{SDW_energy}). Therefore, a phase transition between the normal state and the BEC state is expected at the temperature when  their energy levels cross \cite{Ozawa}.

At intermediate temperature $0<T<T_{\rm c}$, a complete phase diagram is still unclear at present. In principle, the temperature dependence of spin polarization and condensate fraction can be studied via a Hartree-Fock-Bogoliubov approach. However, this problem becomes complicated when the competition between the ZM phase and the SDW phase is considered. We leave this issue to future study.

In summary, a spin-polarized ZM state is identified in the phase diagram of Rashba SO-coupled Bose gases, and the emergence of ZM condensation at finite temperature can be induced by a ferromagnetic transition in the normal state. The phase transition and the critical temperature studied here may be useful to future experiments.

{\it Note Added.}  Recently, K. Riedl and coworkers studied the magnetic transition of the PW phase with $g_{\uparrow\downarrow}<0$ \cite{Riedl}, and a finite condensation temperature was also obtained.

{\it Acknowledgement.}  The author acknowledges helpful discussions with T. Ozawa, A. Recati and L. Yin. This work has been supported by ERC through the QGBE grant and by Provincia Autonoma di Trento.


\begin{thebibliography}{99}
\bibitem{note5} This is a good approximation for current experiments of $^{87}$Rb atoms with Raman coupling \cite{NIST,ChenShuai}, where the difference of intra-spices interactions is within one percent.
\bibitem{note0} We also check the possibility that the condensation wavefunction takes a superposition form of the ZM phase and the SDW phase. Such a BEC state has higher energy.
\bibitem{note4} The dynamical instability occurs for $\delta g\, \bar n<4\epsilon_{k_0}$, where the sound velocity of the ZM phase becomes a imaginary number. However, the ground state of that region is actually either the SDW phase or the PW phase.
\bibitem{note1} In the presence of a Zeeman field $h$, the single-particle spectrum is given by
$\varepsilon_{\bf p,\pm}=\epsilon_{\bf p}\pm\sqrt{h^2+p_\perp^2k_0^2/m^2}$. When $h<2\epsilon_{k_0}$, the low-energy DoS is a constant for $0<\varepsilon-\varepsilon_{\rm min}<(\epsilon_{k_0}-h/2)^2$; when $h>2\epsilon_{k_0}$, the DoS behaves like $D(\varepsilon)\sim \big[1+2\epsilon_{k_0}/(h-2\epsilon_{k_0})\big]\sqrt{\varepsilon-\varepsilon_{\rm min}}$ in the low energy limit.
\bibitem{note2} In the $T\mspace{-3mu}\rightarrow 0$ limit, $\mu\rightarrow\varepsilon_{\rm min}=g\bar n+\frac{1}{2}g_{\uparrow\downarrow}\bar n-\epsilon_{k_0}-\frac{1}{4}(g'\bar nM)^2/\epsilon_{k_0}$, and $\mathcal{F}_{\rm N}\rightarrow N\big[\frac{1}{2}g\bar n+\frac{1}{4}g_{\uparrow\downarrow}\bar n-\epsilon_{k_0} +\frac{1}{4}g'\bar n(2-g'\bar n/\epsilon_{k_0})M^2\big]$. One can see that $M=0$ is favored in this case.
\end{thebibliography}
\end{document}